\documentclass{article}
\usepackage{blindtext}
\usepackage{multicol}
\usepackage{amssymb}
\usepackage{amsmath}
\usepackage{multicol}
\usepackage{graphicx}
\usepackage{float}
\usepackage{caption}
\usepackage{amsmath}
\usepackage{subfig}
\usepackage{epstopdf}
\usepackage{pst-plot}
\usepackage{pgfplots}

\pgfplotsset{width=10cm,compat=1.9}
\usepgfplotslibrary{external}
\restylefloat{table}
\graphicspath{ {./images/} }
\makeatletter \@addtoreset{equation}{section}

\def \be{\begin{equation}}
\def \ee{\end{equation}}
\def \bea{\begin{eqnarray}}
\def \eea{\end{eqnarray}}

\newcommand{\nc}{\newcommand}
\nc{\al}{\alpha} \nc{\bib}{\bibitem} \nc{\la}{\lambda}
\nc{\C}{\mbox{\hspace{1.24mm}\rule{0.2mm}{2.5mm}\hspace{-2.7mm} C}}
\nc{\R}{\mbox{\hspace{.04mm}\rule{0.2mm}{2.8mm}\hspace{-1.5mm} R}}
\graphicspath{ {./Holographic f(Q,T)/} }
\begin{document}

\title{Holographic F(Q,T) gravity with Lambert solution }
\author{Houda Filali$^{1}$,M.Koussour$^{2}$, M. Bennai$^{2}$, Rachid Ahl Laamara$^{1,3}$\thanks{%
Corresponding authors: houda\_filali3@um5.ac.ma \\
pr.mouhssine@gmail.com\\
 mdbennai@yahoo.fr, mohamed.bennai@univh2c.ma \\
 } \\
$^{1}${\small Lab of High Energy Physics, Modeling and Simulations,}\\
\ {\small Faculty of Science, Mohammed V University in Rabat, Morocco}\\
$^{2}${\small Quantum Physics and Applications Team, LPMC, Faculty of
Science Ben M'sik,}\\
\ {\small Casablanca Hassan II University, Morocco}\\
$^{3}${\small CPM, Center of Physics and Mathematics}\\
\ {\small , Faculty of Science, Mohammed V University in Rabat, Rabat,Morocco}}

\maketitle

\begin{abstract}
In this work, we study a model of holographic dark energy using FLRW cosmology in the context of modified gravity. An extension of the symmetric teleparallel gravity is obtained by considering the gravitational action $L$ is given by an arbitrary function $f$ of the non-metricity $Q$, where the nonmetricity Q is responsible for the gravitational interaction, and of the trace of the matter-energy-momentum tensor $T$,so that $L=f(Q,T)$.  We expand on the features of the derived cosmological model in view of the relation between cosmic time and redshift as $t(z)=\frac{kt_{0}}{b}f(z)$ where $f(z)=W\left [ \frac{b}{k}e^{\frac{b-ln(1+z)}{k}} \right ]$ and $W$ denotes the Lambert function, and discuss the evolution trajectories of the equation of state parameter and deceleration parameters in the evolving universe using a special then a generalized version of the model.

\textbf{Keywords:} Modified gravity, holographic dark energy, teleparallel gravity, Lambert solution.

\end{abstract}

\begin{multicols}{2}

\section{\protect\bigskip Introduction}

The accelerated expansion of the universe has been confirmed over the years by several observations from Type Ia supernovae and CMB up to data from multiwavelength observation of blazers and gravitational wave sirens with eLISA\cite{1,2,3,4,5}. This expansion is believed to be the product of dark energy (DE), a mysterious force driven by negative pressure. DE is described as a dark component, and with dark matter (DM), comprises most of the content of our universe today\cite{6}. The exact nature of DE is unknown but many models have been introduced in an attempt to shed light on the mystery and they can be classified into two main categories : 
First, added-on matter components like the cosmological constant $(\Lambda)$ which happens to be the favorite candidate \cite{7}. It originates from vacuum energy and has a constant
EoS of $\omega =-1$, agreeing with observational data. However, the model suffers from various problems such as the fine-tuning and coincidence problem\cite{23,24}.
Other models have been suggested in the same category as an alternative to the cosmological constant, mainly dynamical scalar field models like quintessence, k-essence, or phantom which introduce extra exotic components \cite{8,9,10}. 

The second category includes models that modify the Einstein-Hilbert action by introducing some arbitrary function $f$, changing the geometrical nature of space-time.In this category, we can find various modified gravity theories expanding from Ricci scalar $f(R)$ to Gauss-Bonnet theories.\cite{11,12,29} 
\\
$F(Q)$ gravity provides for an interesting alternative description of spacetime geometry considering that the gravitational interactions described by the non-metricity $Q$ can be presented as an equivalent geometric representation of general relativity without the need for curvature or torsion, relying only on the variation of the length of a vector in the parallel transport in a flat spacetime.The recent interest for this model inspired thorough investigations on both geometrical and physical aspects, from the propagation speed of the gravitational waves to energy conditions and observational constraints, and it does provide with some very interesting results, especially from a cosmological point of view where studies were able to show that the accelerated expansion of the universe was an intrinsic property of the symmetric teleparallel gravity geometry without the need for an exotic dark energy or extra fields.

In this paper, we explore the $f(Q,T)$ gravity\cite{23}, recently introduced as an extension of symmetric teleparallel gravity and represented as an arbitrary function of the non-metricity scalar $Q$ and the trace of matter-energy momentum tensor $T$, where the non-metricity $Q$ is non-minimally coupled to the trace of the matter-energy-momentum tensor $T$, which results in a non-vanishing divergence of the matter-energy-momentum tensor that leads to the appearance of an extra force that can represent either imperfect exotic fluids or quantum phenomena.
The model proved interesting through its compatibility with energy conditions \cite{26} and the non-conservation of the energy-momentum tensor that results from the $Q$ and $T$ coupling yields important thermodynamic changes in the Universe on top of the geometrical ones\cite{28}. 
While this model is able to provide with valid results on the cosmological level by obtaining a late time de Sitter phase that results from an interesting combination of non-metricity and matter, it's not completely able to work as an effective replacement for $\Lambda CDM$ at present times.\cite{23}

In the present work, we attempt to propose a solution for this issue by studying the $f(Q,T)$ model in two different forms, a simplified version and then a more generalized one, assuming the time-redshift relation follows a Lambert function distribution to solve to field equations and find crucial parameters like the energy density, equation of state parameter and deceleration parameter. We introduce Holographic dark energy through an add-on dark energy density with the Hubble horizon as the IR cut-off, which allows us to study the effects of a quantum gravity theory in the framework of teleparallel gravity. This study is especially motivated by recent findings of the Planck telescope confirming the presence of quantum fluctuations during the cosmic inflation era.

This paper is organized in the following manner: Sec. 2  is a general introduction of the $f(Q,T)$ model, Sec. 3 we develop the field equations followed by the subsequent Eos and decelerations parameters; In Sec. 4 we explore the solutions for the field equations by introducing Holographic dark energy and the Lambert function distribution for the redshift. We establish study cases, first a simplified form of the $f(Q, T)$ model then the general form. We analyse the results for each case through the behavior of Eos and deceleration parameter. Sec. 5 is for conclusion and discussion. 
.

\section{ General model}

 Let us consider the extension of the Lagrangian of the symmetric teleparallel gravity, given by the following action \cite{14,26} : 
\begin{equation}
S=\int[\frac{1}{16\pi}f(Q,T)+L_{m}]\sqrt{-g}d^{4}x
\label{a1}
\end{equation}%

Where $f(Q,T)$ is a general function of $Q$ and the trace of the matter-energy-momentum tensor $T$, $L_{m}$ is the matter Lagrangian and  $g\equiv det(g_{\mu \nu })$. 
\newline
\newline
The non-metricity tensor is defined as 
\begin{equation}Q_{\gamma\mu\nu}=\nabla_{\gamma}g_{\mu\nu}\end{equation}
 and traces of the non metricity tensor can be derived  as the following :  

\begin{equation} Q_{\delta}=g^{\mu\nu}Q_{\delta\mu\nu} , \tilde{Q}_{\delta}=g^{\mu\nu}Q_{\mu\delta\nu}\end{equation}
We also define the Super potential or non-metricity conjugate as : 
\begin{equation} P^{\delta}_{\mu\nu}=-\frac{1}{2}Q^{\delta}_{\mu\nu}+\frac{1}{4}(Q^{\delta}-\tilde{Q}^{\delta})g_{\mu\nu}-\frac{1}{4}\delta^{\delta}(\mu Q\nu) \end{equation}
Where the trace of the  nonmetricity tensor becomes : 
\begin{equation}Q=-Q_{\delta\mu\nu}P^{\delta\mu\nu}\end{equation}
The energy-momentum tensor is defined as : \begin{equation}T_{\mu\nu}=\frac{-2}{\sqrt{-g}}\frac{\delta\sqrt{-g}L_{m}}{\delta g^{\mu\nu}}\end{equation}
\begin{equation} \theta_{\mu\nu}=g^{\alpha\beta}\frac{\delta T_{\alpha\beta}}{\delta g^{\mu\nu}}\end{equation}
Variation of $T_{\mu\nu}$ w.r.t $g_{\mu\nu}$ leads to the following equation : \begin{equation}\frac{\delta g^{\mu\nu}T_{\mu\nu}}{\delta g^{\alpha\beta}}=T_{\mu\nu}+\theta_{\mu\nu}\end{equation}
\newline
Variation of action (2.1) with respect to the metric tensor components yields the field equations for the symmetric teleparallel  $f(Q, T)$ gravity as :
\begin{equation}
\begin{gathered}
-\frac{2}{\sqrt{-g}}\nabla_{\delta}(f_{Q}\sqrt{-g}P^{\delta}_{\mu\nu})-\frac{1}{2}fg_{\mu\nu}+f_{T}(T_{\mu\nu}+\theta_{\mu\nu}) \\ -f_{Q}(P_{\mu\delta\alpha}Q^{\delta\alpha}_{\nu}-2Q^{\delta\alpha}_{\mu}P_{\delta\alpha\nu}) =8\pi T_{\mu\nu}
\end{gathered}
\end{equation}

where $f_{Q}=\frac{df}{dQ}$, $f_{T}=\frac{df}{dT}$ and $\nabla_{\delta}$ is the covariant derivative.

\section{\protect Flat FLRW Universe in $f(Q,T)$ }

We now consider a flat, homogenous and isotropic FLRW spacetime of the form :
\begin{equation}ds^2=-N(t)^2dt^2+a(t)^2(dx^2+dy^2+dz^2) \end{equation}
 where a(t) represents the scale factor and the lapse function N(t)=1 for a flat spacetime.
\newline
With the Hubble parameter being  $H\equiv \frac{\dot{a}}{a}$ , the non metricity tensor becomes :  \begin{equation}  Q=6\frac{H^2}{N^2}\Rightarrow Q=6H^2\end{equation}
Considering a perfect fluid, we get  : 
\begin{align*}T_{\nu}^{\mu}=diag(-\rho ,p,p,p)\\
 \theta _{\nu}^{\mu}=diag(2\rho+p ,-p,-p,-p)
\end{align*}
we finally obtain the modified Friedman equations with N=1 as
 \begin {equation}
8\pi \rho =\frac{f}{2}-6FH^{2}-\frac{2\tilde{G}}{1+\tilde{G}}(\dot{F}H+F\dot{H}) \end{equation}
and
\begin{equation}8\pi p= -\frac{f}{2}+6FH^{2}+2(\dot{F}H+F\dot{H})
 \end{equation}
where dot represents derivative with respect to time,  $F=f_{Q}$ and $8\pi\tilde{G}=f_{T}$.
\\
Moreover, it is interesting to mention that the divergence of the matter-energy-momentum tensor of the $F(Q,T)$ gravity can be expressed as :
\begin{eqnarray}
\hspace{-0.5cm} &&\mathcal{D}_{\mu }T_{\ \ \nu }^{\mu }=\frac{1}{f_{T}-8\pi }%
\Bigg[-\mathcal{D}_{\mu }\left( f_{T}\Theta _{\ \ \nu }^{\mu }\right) -\frac{%
16\pi }{\sqrt{-g}}\nabla _{\alpha }\nabla _{\mu }H_{\nu }^{\ \ \alpha \mu } 
\notag \\
\hspace{-0.5cm} &&+8\pi \nabla _{\mu }\bigg(\frac{1}{\sqrt{-g}}\nabla
_{\alpha }H_{\nu }^{\ \ \alpha \mu }\bigg)-2\nabla _{\mu }A_{\ \ \nu }^{\mu
}+\frac{1}{2}f_{T}\partial _{\nu }T\Bigg],
\end{eqnarray}%
where $H_{\gamma }^{\ \ \mu \nu }$ is the hyper-momentum tensor density
defined as,%
\begin{equation}
H_{\gamma }^{\ \ \mu \nu }\equiv \frac{\sqrt{-g}}{16\pi }f_{T}\frac{\delta T%
}{\delta \widetilde{\Gamma }_{\ \ \mu \nu }^{\gamma }}+\frac{\delta \sqrt{-g}%
\mathcal{L}_{M}}{\delta \widetilde{\Gamma }_{\ \ \mu \nu }^{\gamma }}.
\end{equation}
The above equation shows that in the $f(Q,T)$ gravity theory,
the matter-energy-momentum tensor is not conserved, i.e. $\hspace{-0.1cm}%
\mathcal{D}_{\mu }T_{\ \ \nu }^{\mu }\neq 0$. This divergence can be
interpreted as an additional force or exotic field. It also indicates the amount of energy that either enters or exits a specific volume of a physical system.
 It is worth mentioning that in the absence of the $f(T)$ term we recover the results of the $f(Q)$ model and the energy-momentum tensor becomes conserved
\\
The evolution equation for the Hubble function can then be obtained in the following form : 
\begin{equation}
\dot{H}+\frac{\dot{F}}{F}H=\frac{4\pi }{F}(1+\tilde{G})(\rho +p)
\end{equation}
And the equivalent Friedmann equations are ; 
\begin{equation} 8\pi \rho _{eff}=3H^{2}=\frac{f}{4F}-\frac{4\pi}{F}((1+\tilde{G})\rho +\tilde{G}_{p})
\end{equation}
and
 \begin{equation}
\begin{gathered}
-8\pi\rho_{eff}=2\dot{H}+3H^{2} \\ =\frac{f}{4F}-\frac{2\dot{F}H}{F}+\frac{4\pi}{F}((1+\tilde{G})\rho +(2+\tilde{G})p)
\end{gathered}
\end{equation}

 Using the equations for $\rho$ and $p$ from (3.8) and (3.9) in the framework of our $f(Q,T)$ model and from eqs. (3.4), (3.5) and (3.6) we can write the density and pressure of our modified gravity model in the form : 
\begin{equation} \rho = \frac{1}{8\pi }[\frac{f}{2}-6FH^{2}-\frac{2\tilde{G}}{1+G}\chi ]\end{equation}
and
\begin{equation}p = \frac{1}{8\pi }[-\frac{f}{2}+6FH^{2}+2\chi ]\end{equation}
with $\chi = \dot{F}H+F\dot{H}$
\newline

 The expression for the Equation of State parameter $(\omega = \frac{p}{\rho})$ , which is a function of the energy density and pressure, is an important parameter for the classification of the different stages of the Univers's expansion. The late-time accelerated expansion is represented by $\omega < -\frac{1}{3}$ with varying models existing within this range. The EoS parameter for our $f(Q,T)$ model can be written : 
 \begin{equation}\omega =-1+\frac{4(\dot{F}H+F\dot{H})}{(1+\tilde{G})(f-12Fh^{2})-4\tilde{G}(\dot{F}H+F\dot{H})}
\end{equation}
And  deceleration parameter : 
\begin{equation}
q=-1+\frac{\dot{F}H+4\pi(1+\tilde{G})(\rho +p)}{FH^{2}}\end{equation}
To compare our model with astronomical observations and established cosmological models, we introduce the redshift $z$, instead of the time variable $t$, defined as :
\begin{equation}
    1+z=\frac{1}{a}
\end{equation}
The decelaration parameter as a function of $z$ can be obtained as :
\begin{equation}
    q(z)=(1+z)\frac{1}{H(z)}\frac{dH(z)}{dz}-1
\end{equation}
We assume the late universe is filled with dust matter only, on negligible pressure. We can then obtain the standard general relativistic energy conservation equation $\dot{\rho}+3H\rho=0$, which allows us to write the evolution of the Hubble function in the following form : 
\begin{equation}
    H=H_{0}\sqrt{(\Omega_{DM}+\Omega_{b})a^{-3}+\Omega_{\Lambda}}
\end{equation}
where $\Omega_{DM}$,$\Omega_{b}$ and $\Omega_{\Lambda}$ are the density parameters of the cold dark matter, baryonic matter, and dark energy (interpreted as a cosmological constant), respectively. The density parameters satisfy the important constraint $\Omega_{DM}$ + $\Omega_{b}$ + $\Omega_{\Lambda}$ = 1.
The deceleration parameter as a function of the redshift $z$ becomes \cite{23} : 
\begin{equation}
    q(z)=\frac{3(1+z)^{3}(\Omega_{DM}+\Omega_{b})}{2[\Omega_{\Lambda}(1+z)^{3}(\Omega_{DM}+\Omega_{b})]}
\end{equation}
We now introduce the jerk parameter $(j)$ as a higher derivative of the deceleration $q$, in the following form \cite{34,35}: 
\begin{equation}
    j = (1+z)\frac{dq}{dz}+q(1+2q)
\end{equation}
From equation (3.17) and (3.18), the jerk parameter becomes : 
\begin{eqnarray}
  \hspace{-0.5cm} &&\frac{3}{2}\Bigg[\frac{(1+z)^{2}(\Omega_{DM}+\Omega_{b})[3(\Omega_{\Lambda}+(1+z)^{3}(\Omega_{DM}+\Omega_{b}))}{[\Omega_{\Lambda}+(1+z)^{3}(\Omega_{DM}+\Omega_{b})]^{2}}
  \notag \\
  \hspace{-0.5cm} && -\frac{3(1+z)^{3}(\Omega_{DM}+\Omega_{b})]}{[\Omega_{\Lambda}+(1+z)^{3}(\Omega_{DM}+\Omega_{b})]^{2}}\Bigg]
\end{eqnarray}
The sign of jerk parameter $j$ regulates the change of the Universe’s
dynamics, a positive value indicating the instance of a transition time under which the Universe modifies the expansion. In the following for the density parameters we adopt the numerical values $\Omega_{DM}$ = 0.2589, $\Omega_{DM}$= 0.0486, and $\Omega_{DM}$ = 0.6911. From these numerical values of the cosmological parameters, it follows that the present value of the jerk parameter $j(0)= 0.95$. While we do not obtain the usual $j_{\Lambda}=1$ of the $\Lambda CDM$ model, our results remain very close to it. 

\section{Holographic dark energy, Lambert function and cosmological solution}
\subsection{The Holographic dark energy model with Hubble cut-off}
As one of the most promising dynamical dark energy models of the recent past, the holographic dark energy model has laid an interesting template for research that aims to navigate the dark energy problem. Considering the holographic principle  \cite{27} and Bekenstein bound \cite{21}, the vacuum energy of a region of size L should not cross the bound of the mass of a black hole of the same size. In the cosmological context, the holographic principle sets an upper bound on the entropy of the universe. According to Cohen et al.\cite{22}, to maintain effective quantum field theory, this bound in a region of size L with a short cut-off should take the following form : 
\begin{equation}L^{3}\Lambda ^{3}\leq S_{BH}=\pi L^{2}M_{p}^{2}\end{equation}
Where $\Lambda$ is the UV cut-off, $S_{BH}$ is the entropy of a black hole of size L and $M_{p}$ is the reduced Planck mass. 
\newline
Converting this relation to having the entire energy of a region of size L not exceed the mass of a black hole of the same size, gives us the equation for the density of Holographic dark energy bound : 
\begin{equation} \rho _{\Lambda }\leq L^{-2}M_{p}^{2}\end{equation}

We choose to work in the case of the highest bound, giving us $\rho_{\Lambda}=3c^{2}M_{p}^{2}L^{-2}$, with $c$ being a free dimensionless parameter and we consider $c^{2}=M_{p}^{2}=1$ for the rest of our work.
The IR cut-off is supposed to be the particle horizon $L_{p}$ or the future event horizon $L_{F}$ , which are determined respectively as :
\begin{equation}
    L_{p} \equiv a \int_0^t \frac{dt}{a} ,\hspace{0.5cm} L_{F} \equiv a \int_t^\infty \frac{dt}{a}
\end{equation}
Taking the time derivative of the above equations, we get the expressions for the Hubble parameter in terms of  $L_{p}$ , $L_{F}$ , and their time derivatives as follows :
\begin{equation}
    H( L_{p}, \dot{ L_{p}}) \equiv \frac{\dot{ L_{p}}}{ L_{p}}-\frac{1}{ L_{p}} , \hspace{0.5cm}H( L_{F}, \dot{ L_{F}}) \equiv \frac{\dot{ L_{F}}}{ L_{F}}+\frac{1}{ L_{F}}
\end{equation}

In the generalized form of the Holographic dark energy  proposed by Nojiri \cite{30, 31} the infrared cutoff is presented with a combination of the natural FRW parameters such as Hubble rate, particle and future horizons :

\begin{equation}
    L_{IR}=L_{IR}(L_{p},\dot{L_{p}},\Ddot{L_{p}},....,L_{F},\dot{L_{F}},\Ddot{L_{F}},...a)
\end{equation}
As a more  general form than (4.4), we may consider the case that  $L_{IR}$ depends on the Hubble rate $H$ and the length scale $l$ coming from the cosmological constant $\Lambda = 12/l^{2}$ as shown in \cite{30} :
\begin{equation}
    L_{IR}=L_{IR}(L_{p},L_{F},t_{s},H,l)
\end{equation}
Where $L_{IR}$ can take the following form : 
\begin{eqnarray}
  \hspace{-0.5cm} &&\frac{c}{L_{IR}}= \frac{H}{\alpha+1}+\frac{1}{L_{F}}\Bigg[1+\frac{2(\alpha^{2}-\alpha-1}{\alpha +1})\frac{L_{F}}{\alpha l} 
  \notag \\
 \hspace{-0.5cm} && + (1-\frac{\alpha^{2}(\alpha + 2)}{\alpha + 1}(\frac{L_{F}}{\alpha l}^{2}\Bigg]
\end{eqnarray}
Where $\alpha$ is a positive dimensionless parameter. 
In the present paper, considering the choice of our model which is non minimally coupled to the trace of the matter energy-momentum tensor $T$, we choose to work with the special case $\alpha = 0$, therefore retrieving $\frac{c}{L_{IR}} = H$ with $c=1$ and the Hubble parameter as the horizon cut off :
\begin{equation} L_{IR} = H^{-1}\end{equation}
\begin{equation} \rho_{DE}=3L^{-2}=3H^{2}\end{equation}
\subsection{Lambert function solution and results}

If we attempt to solve the equations (3.7), (3.8) and (3.10) , it would be very hard to do so without any kind of further simplification considering that we have several unknown parameters and variables. We already replaced the expression for the density of dark energy by introducing the Holographic model and choosing the Hubble constant at the IR cut-off, but the rest of the equations will require a mathematical approach to make them solvable. 
\newline

For this reason, we assume that the redshift dependant time equation takes the form of a Lambert function distribution, as : 
 
\begin{equation}t(z)=\frac{mt_{0}}{l}g(z)\end{equation}
With $g(z)=LambertW(\frac{l}{m}e^{\frac{l-ln(1+z)}{m}})$, $m$ and $l$ are non negative constants and $t_{0}$ represents the age of the universe.

By using the Lambert solution, we can rewrite the previous equations as follows : 
\newline
Hubble function : 
\begin{equation} H=\frac{m}{t}+\frac{l}{t_{0}}\end{equation}
Holographic dark energy density : 
\begin{equation}\rho_{DE}=3[\frac{m}{t}+\frac{l}{t_{0}}]^{2}\end{equation}
We chose the model parameters as $\alpha = -7$ and $\beta = -1$ to make the energy density positive and the EoS parameter as per the observations.

\begin{figure}[H]
\begin{minipage}[c]{1\linewidth}
\includegraphics[width=\linewidth]{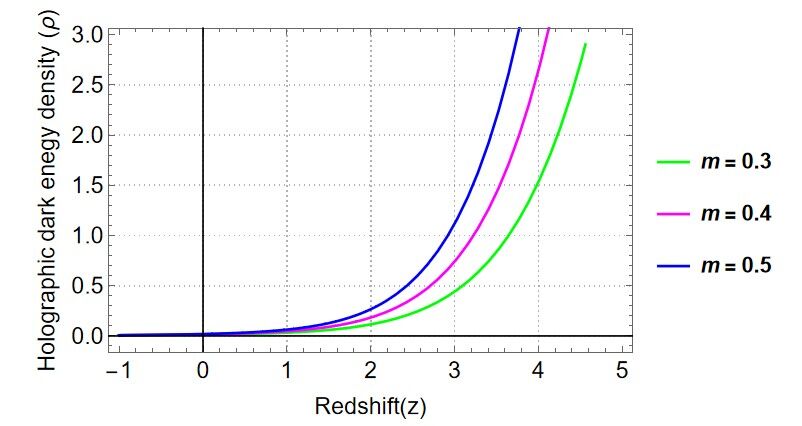}
\caption{Density of Holographic dark energy with different values of $m$.}
\end{minipage}
\end{figure}

We can consider any of the following forms of $f(Q,T)$ : 
\begin{equation}\left\{\begin{matrix}
f(Q,T)=\alpha Q+\beta T &  & \\ 
f(Q,T)=\alpha Q^{n+1}+\beta T &  & \\ 
f(Q,T)=-\alpha Q-\beta T^2 &  & 
\end{matrix}\right. \end{equation}
For the purpose of simplicity, we choose to work first with the case $f(Q,T)=\alpha Q+\beta T$,  where $\alpha$ and $\beta$ are model parameters. The values of both $f_{Q}$ and $f_{T}$ are obtained as $f_{Q}= F = \alpha$ and $f_{T} = 8\pi G = \beta$.

The equation of state parameter of Holographic dark energy becomes : 
\begin{equation} \omega _{DE}=\frac{3\alpha( \frac{l}{t_{0}}-\frac{m}{t})^{2}-\alpha (2+G-\frac{G^{2}}{1+G})\frac{m}{t^{2}}}{6\pi (\frac{m}{t}+\frac{l}{t_{0}})^{2}[(2+G)^{2}+4G-G^{2}]}\end{equation}

\begin{figure}[H]
\begin{minipage}[c]{1\linewidth}
\includegraphics[width=\linewidth]{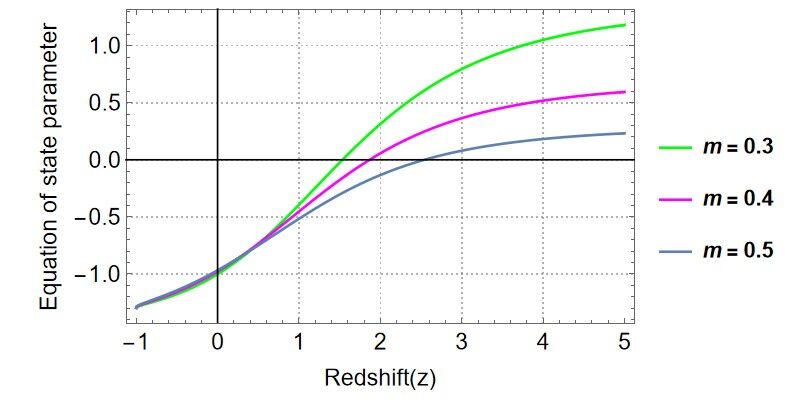}
\caption{Equation of state parameter of holographic dark energy in Hubble’s cut-off of the $f(Q,T)=\alpha Q+\beta T$ gravity model.}
\end{minipage}
\end{figure}

In Figures 1 and 2 we plotted the behaviors of the density of Holographic dark energy $\rho_{DE}$ and the equation of state parameter $\omega_{DE}$ with the Hubble horizon cut-off in term of redshift (z) for the three different values of $m=0.3$, $m=0.4$ and $m=0.5$.
We notice that both functions are increasing with redshift, with the Eos function going from negative to positive values around $z = 1.5$. We can see how the variations of the function represent the cosmological evolution of the universe going from a quintessence state $(-1 <\omega<-0.33)$  to a phantom phase $(\omega < -1)$, crossing the $\Lambda CDM$ state with $\omega = -1$ at present times, as observed in astronomical data, with that being correct for all values of $m$.
 While values of the Holographic dark energy density are strictly positive, including in present time, which reflects active action of quantum fluctuations still at work.

We attempt the same procedure for the general form of our$f(Q,T)$ model : $f(Q,T)=\alpha Q^{n+1}+\beta T $ , which results in the following EoS parameter : 
\begin{equation}
\omega ' = \frac{-2\pi(1+G)(\frac{m}{t}+\frac{t}{t_{0}}^2)+\alpha(2n+1)(n+1)\frac{-m}{t^2}(\frac{m}{t}+\frac{l}{t_{0}})^{2n}}{2\pi(\frac{m}{t}+\frac{l}{t_{0}})^2(1+G)}
\end{equation}

\begin{figure}[H]
\begin{minipage}[c]{1\linewidth}
\includegraphics[width=\linewidth]{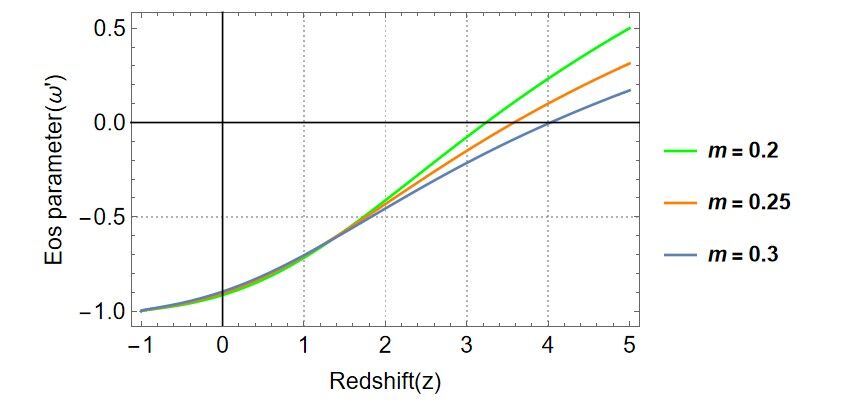}
\caption{Equation of state parameter of holographic dark energy in Hubble’s cut-off of the generalized $f(Q,T)$ gravity model .}
\end{minipage}
\end{figure}

In Figure 3, we can see how the behavior of the EoS parameter changes when we study the general form of our teleparallel gravity model. The choice of constants stays generally unchanged except for the introduction of a new constant ( $n=0.1$) for this form of the model. We notice that the curve still represents an evolution from a decelerated to an accelerated expansion, especially at late time, the transition being around $z=2.2$. But contrary to the previous simplified $f(Q,T)$, we do not find $\omega = -1$ at present time but rather a model that behaves like quintessence for the different values of $m$ and only reaches the $\Lambda CDM$ constant later in time, avoiding a phantom scenario all together.

Figure 4 represents the variation of the deceleration parameter $q$ with respect to the redshift (z). We can see how values of the parameter go from negative to positive around $z=0.9$ , depicting the transition from deceleration to the current acceleration phase. These results show that our model is consistent with recent observational findings. 
\newline
\newline

\begin{figure}[H]
\begin{minipage}[c]{1\linewidth}
\includegraphics[width=\linewidth]{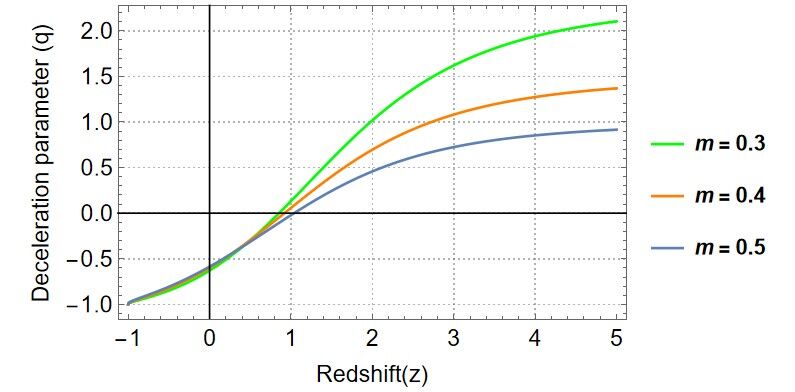}
\caption{Deceleration parameter.}
\end{minipage}
\end{figure}

\section{Conclusion}

In this paper, we have investigated the Holographic dark energy in the framework of the asymmetric Teleparallel gravity model $f(Q,T)$, where the non-metricity scalar $Q$ is responsible for the gravitational interaction and $T$ is the trace of the matter-energy-momentum tensor. We considered two different forms of our model. First, a simplified version represented by $f(Q,T) = \alpha Q +\beta T$ then a more general form represented by : $f(Q,T) = \alpha Q^{n+1}+\beta T$  where $\alpha$, $\beta$ and $n$ are free model parameters. In addition, we assumed that the redshift-time relation follows the form of a Lambert function distribution for simplified results.

 We study the behavior of various cosmological parameters such as the energy density, Equation of state parameter, and deceleration parameter from which we get the following results: We see that the Holographic dark energy density is a decreasing function of redshift $z$ that evolves in positive values for all choices of the constant $m$, especially in present time $(z = 0)$ where it represents the non-vanishing of quantum fluctuations (Fig. 1) and a clear influence on the accelerated late time expansion.
We study the evolution of the deceleration parameter with respect to the redshift (z) and we can observe a transition from decelerated to accelerated expansion around $(z=1)$ for all values of $m$.

 Furthermore, we observed the evolution of the EoS parameter, which is a function of the holographic dark energy density and pressure in the framework of our $f(Q, T)$ model, and it shows the transition from the matter-dominated era to the accelerated expansion, first through quintessence then into the $\Lambda CDM$ value of $\omega = -1$ at present time, which is in agreement with all observational data, and finally evolving into phantom-like behavior in later time. This is correct for all chosen values of $m$ which are perfectly compatible at present and later time but show differences in the redshift at which the model enters the accelerated expansion phase with $m=0.5$ being the earliest $(z=2.3)$ and $m=0.3$ the latest $(z=1.2)$. We see how the introduction of Holographic dark energy is able to provide better cosmological results within the $F(Q,T)$ gravity model at present times that are compatible with observational data and can therefor provide a good description of our universe and plausible predictions for it's future and eventually be a proper alternative to the $\Lambda CDM$ model, while also being a plausible candidate for the unknown force that is generated from non conservation of the matter-energy-momentum tensor which we mentioned can be interpreted by quantum phenomena.

The generalized form of our model shows some different results. First, the accelerated expansion phase starts at a later redshift for all values of $m$, resulting in the study of dark energy being independent of the free parameter we choose for it. Furthermore, the function for this model stays in a quintessence state for much longer, even at present time, and only reaches the $\omega = -1$ value at a later time, avoiding a phantom state scenario.
the results for the first form the of $f(Q,T)$ model, which are in agreement with astronomical data, are a special case. While the results for the generalized model do not concord with the current data, they do however present a theoretically viable model that presents stable future predictions while also respecting the various energy conditions. \cite{26}

The main difference that our model proposes is the study of HDE in the complete absence of curvature which is an interesting angle considering that HDE finds its source in black hole entropy. While our present paper focuses on the study of late-time acceleration, the premise of our model opens the door for a redefinition of black hole dynamics and even a new form of early time inflation
Another possible application of the $f(Q, T)$ theory is to consider inflation in the presence of scalar fields, an approach that provides useful insights for the description of the early phases of cosmological evolution. and therefore a new unified model can be explored in future work.

 We can conclude that the presented approach predicts de Sitter type expansions of the Universe, and thus it may represent a geometric alternative to dark energy. Furthermore, our specific model of Holographic $F(Q,T)$ gravity can produce results that are very similar to the $\Lambda CDM$ model when compared to recent observational data, with an all time positive value of the density of dark energy, an EoS parameter that goes from matter dominated era to a present day value of $\omega = -1$, a deceleration parameter of $q(z=0)=-0.6$ and $q(z=-1)=-1$ and a jerk parameter of $j = 0.95$. All these cosmological bounds, in addition to energy conditions and observational constraints that have previously been explored , show that our model can be considered as a serious alternative to GR, especially for late time acceleration.

\end{multicols}


\begin{thebibliography}{99}
\bibitem{1}A. G. Riess et al., Observational evidence from supernovae for an accelerating Universe and a cosmological
constant, The Astronomical Journal 116 (1998) 1009.

\bibitem{2}S. Hanany et al., MAXIMA-1: a measurement of the cosmic microwave background anisotropy on angular scales
of 10’-5, The Astrophysical Journal 545 (2000) L5.

\bibitem{3}D. J. Eisenstein et al., Detection of the baryon acoustic
peak in the large-scale correlation function of SDSS luminous red galaxies. The Astrophysical Journal 633 (2005)
560.
\bibitem{4}Domínguez, A.,  Prada, F. (2013). Measurement of the Expansion Rate of the Universe from delta-Ray Attenuation. The Astrophysical Journal Letters, 771(2), L34.

\bibitem{5}Tamanini, N., Caprini, C., Barausse, E., Sesana, A., Klein, A.,  Petiteau, A. (2016). Science with the space-based interferometer eLISA. III: Probing the expansion of the Universe using gravitational wave standard sirens. Journal of Cosmology and Astroparticle Physics, 2016(04), 002.

\bibitem{6}Bennett, C. L., Bay, M., Halpern, M., Hinshaw, G., Jackson, C., Jarosik, N., ...  Wright, E. L. (2003). The microwave anisotropy probe* mission. The Astrophysical Journal, 583(1), 1.

\bibitem{7}Peebles, P. J. E.,  Ratra, B. (2003). The cosmological constant and dark energy. Reviews of modern physics, 75(2), 559.

\bibitem{8}P. Armendariz, V. M. Christian, Mukhanov, and J. S. Paul. Essentials of k-essence, Physical Review D 63 (2001)103510.

\bibitem{9}Jamil, M., Myrzakulov, Y., Razina, O.,  Myrzakulov, R. (2011). Modified Chaplygin gas and solvable F-essence cosmologies. Astrophysics and Space Science, 336, 315-325.

\bibitem{10}Nojiri, S. I.,  Odintsov, S. D. (2003). Quantum de Sitter cosmology and phantom matter. Physics Letters B, 562(3-4), 147-152.

\bibitem{11}Nojiri, S. I.,  Odintsov, S. D. (2011). Unified cosmic history in modified gravity: from F (R) theory to Lorentz non-invariant models. Physics Reports, 505(2-4), 59-144.

\bibitem{12}Nojiri, S. I.,  Odintsov, S. D. (2007). Introduction to modified gravity and gravitational alternative for dark energy. International Journal of Geometric Methods in Modern Physics, 4(01), 115-145.

\bibitem{13}T. Harko et al. f(R, T) gravity. Physical Review D 84 (2011)
024020


\bibitem{14} Bhattacharjee, S., Sahoo, P.K. Baryogenesis in f(Q,T)
 gravity. Eur. Phys. J. C 80, 289 (2020).
\bibitem{15}Shekh, S. H. (2021). Models of holographic dark energy in f (Q) gravity. Physics of the dark Universe, 33, 100850.
\bibitem{16}Koussour, M., Shekh, S. H., Filali, H.,  Bennai, M. (2023). Barrow holographic dark energy models in f (Q) symmetric teleparallel gravity with Lambert function distribution. International Journal of Geometric Methods in Modern Physics, 20(02), 2350019.
\bibitem{17}Shekh, S. H., Moraes, P. H.,  Sahoo, P. K. (2021). Physical acceptability of the Renyi, tsallis, and Sharma-Mittal holographic dark energy models in the f (t, b) gravity under Hubble’s cutoff. Universe, 7(3), 67.
\bibitem{18}Adak, M. (2006). The symmetric teleparallel gravity. Turkish Journal of Physics, 30(5), 379-390.

\bibitem{19}Wang, S., Wang, Y.,  Li, M. (2017). Holographic dark energy. Physics reports, 696, 1-57.
\bibitem{20}Li, M. (2004). A model of holographic dark energy. Physics Letters B, 603(1-2), 1-5.
\bibitem{21}BekensteinJ.D.. Black holes and entropy. Phys. Rev. D (1973)
\bibitem{22}Hooft, G. T. (1985). On the quantum structure of a black hole. Nuclear Physics B, 256, 727-745.
\bibitem{23}Xu, Y., Li, G., Harko, T.,  Liang, S. D. (2019). f (Q, T) gravity. The European Physical Journal C, 79, 1-19.
\bibitem{24}Das, S.,  Mandal, S. (2023). Aspects of Cosmology in Symmetric Teleparallel f (Q, T)  Gravity. arXiv preprint arXiv:2306.14912.
\bibitem{25}El Bourakadi, K., Koussour, M., Otalora, G., Bennai, M.,  Ouali, T. (2023). Constant-roll and primordial black holes in f (Q, T) gravity. Physics of the Dark Universe, 41, 101246.
\bibitem{26}Arora, S.,  Sahoo, P. K. (2020). Energy conditions in f (Q, T) gravity. Physica Scripta, 95(9), 095003.
\bibitem{27}Fischler, W.,  Susskind, L. (1998). Holography and cosmology. arXiv preprint hep-th/9806039.
\bibitem{28}Harko, T.,  Lobo, F. S. (2018). Extensions of f (R) Gravity: Curvature-Matter Couplings and Hybrid Metric-Palatini Theory (Vol. 1). Cambridge University Press.
\bibitem{29}Koussour, M., Filali, H., Shekh, S. H.,  Bennai, M. (2022). Holographic dark energy in Gauss-Bonnet gravity with Granda-Oliveros cut-off. Nuclear Physics B, 978, 115738.
\bibitem{30}Nojiri, S. I.,  Odintsov, S. D. (2006). Unifying phantom inflation with late-time acceleration: Scalar phantom–non-phantom transition model and generalized holographic dark energy. General Relativity and Gravitation, 38, 1285-1304.
\bibitem{31}Nojiri, S. I., Odintsov, S. D.,  Paul, T. (2021). Different faces of generalized holographic dark energy. Symmetry, 13(6), 928.
\bibitem{32}Myrzakulov, N., Koussour, M., Alfedeel, A. H.,  Elkhair, H. M. (2023). Cosmological implications of the constant jerk parameter in f (Q, T) gravity theory. Chinese Journal of Physics, 86, 300-312.
\bibitem{33}Arora, S., Parida, A.,  Sahoo, P. K. (2021). Constraining effective equation of state in f (Q, T) gravity. The European Physical Journal C, 81, 1-7.
\bibitem{34}S. Pan, A. Mukherjee, N. Banerjee, Mon. Not. R. Astron.Soc. 477, 1189 (2018)
\bibitem{35} S. Mandal, S. Bhattacharjee, S. K. J. Pacif, P.K. Sahoo, Phys.Dark Univ. 28, 100551 (2020).
\end{thebibliography}
\end{document}